# LOCAL RESONANCE AND BRAGG BANDGAPS IN SANDWICH BEAMS CONTAINING PERIODICALLY INSERTED RESONATORS

B. Sharma[a]* and C.T. Sun[b]

[a] School of Industrial Engineering, Purdue University, West Lafayette, Indiana, 47907, USA

[b] School of Aeronautics and Astronautics, Purdue University, West Lafayette, Indiana, 47907, USA

**ABSTRACT**

We study the low frequency wave propagation behavior of sandwich beams containing periodically embedded internal resonators. A closed form expression for the propagation constant is obtained using a phased array approach and verified using finite element simulations. We show that local resonance and Bragg bandgaps coexist in such a system and that the width of both bandgaps is a function of resonator parameters as well as their periodicity. The interaction between the two bandgaps is studied by varying the local resonance frequency. We find that a single combined bandgap does not exist for this system and that the Bragg bandgaps transition into sub-wavelength bandgaps when the local resonance frequency is above their associated classical Bragg frequency.

* Corresponding author. School of Industrial Engineering, Purdue University, West Lafayette, Indiana, 47907, USA Tel.: +1 765 404 0565.

*E-mail address*: bsharma@purdue.edu (B. Sharma).

**INTRODUCTION:**

The creation of wave attenuation frequency bands due to addition of substructures to a host medium was first demonstrated in the context of acoustic waves by Liu et al. [1]. Various researchers have since studied the attenuation of acoustic and elastic waves by utilizing such locally resonating elements [2]. The localized resonance of the added substructures causes energy sequestering and prohibits the propagation of incident waves [3]. These local resonance bandgaps are distinct from those obtained due to multiple scattering and interference effects occurring between periodic elements in a propagation medium, classically referred to as Bragg bandgaps [4], [5], and [6]. Due to the nature of the mechanism involved in their creation, Bragg bandgaps are generated at wavelengths comparable to the spatial scale of the periodicities [5]. Local resonance bandgaps, on the other hand, are independent of the spatial organization of the resonant substructures and are solely governed by the unit cell resonance frequency [7].

Recently, researchers have investigated various systems containing periodically placed local resonators and demonstrated the coexistence of both bandgaps in the same system [8 - 20]. Still et al. [8] experimentally demonstrated the existence of hypersonic Bragg as well as local resonance bandgaps in three dimensional colloidal films of nanospheres and also showed that for a structurally disordered system, the Bragg bandgap disappears while the local resonance bandgap persists. Croënne et al. [9] reported their coexistence as well as an overlap between the two bandgaps for a 2D crystal of nylon rods in water. Similarly, Achaoui et al. [10] showed the presence of local resonance bandgaps at low frequencies and Bragg bandgaps at high frequencies for surface guided waves in a lithium niobate substrate with nickel pillars, while Bretagne et al. [11] reported similar results for acoustic wave propagation through a 3D bubble phononic crystal. Kaina et al. [12], Chen et al. [13], and Yuan et al. [14] have recently reported results demonstrating coupling between the two bandgaps and creation of a single combined 'resonant-Bragg' bandgap extending over a wide frequency range. For flexural wave propagation, Liu et al [15] utilized a transfer matrix method to study the effect of various periodicities and their combination on the bandgap characteristics of a beam. Specifically, for the case of suspended mass periodicity they showed that local resonance and Bragg bandgaps coexist in the system and also provided a transition criterion for the two bandgaps in terms of the group velocity gradient. Others have used a similar transfer matrix approach to study different beam structures with periodically placed resonators [16], [17], [18]. Xiao et al used the plane wave expansion method [19] and a spectral element formulation [20] to obtain the propagation constant for an Euler-Bernoulli beam with attached local resonators, and demonstrated a super-wide combined pseudo gap when a resonance gap was nearly coupled with a Bragg bandgap.

The idea of the utilization of space available in thick sandwich cores to accommodate locally resonating elements was first proposed by Chen et al. [21]. The high stiffness to weight ratio of sandwich structures makes them an ideal solution to save energy through weight reduction. Consequently, sandwich beams have gradually found increased applications in the aerospace, marine as well as the automotive sectors [22]. However, a major roadblock to further adoption of composite sandwich designs is their susceptibility under dynamic loading [23]. Chen et al. [24] demonstrated that embedding resonators inside the sandwich core generates local resonance wave attenuation bandgaps which help improve their dynamic flexural performance without a significant weight penalty. To analyze the effect of resonators embedded inside the core, they assumed the resonators to be uniformly distributed over the entire length of the beam using a volume averaging technique. Doing so allows the description of such a resonator-embedded beam using continuous field variables and the equations of motion were derived using Hamilton's principle. However, this model did not account for discrete resonators inserted in the core with a spatial periodicity and was unable to capture the effect of such a periodicity on the dispersion behavior of the system.

In this study, we adopt the phased array method to obtain the propagation constants for a sandwich beam with resonator embedded core. The phased array method was developed by Mead [25] to obtain closed form solutions for the propagation constants of various periodic systems. Since we are primarily interested in the low frequency behavior of the system, we model the sandwich beam as an equivalent Timoshenko beam [21], [26] and treat the resonators as a phased array of forces. Dispersion curves obtained by this method are compared with volume averaging method and finite element results and it is shown that local resonance and Bragg bandgaps coexist. The relationship between bandgap bounding frequencies, resonator stiffness and mass, and the periodic distance is analyzed in the context of modal frequencies of simple unit cell models [6]. Finally, the interaction between local resonance and Bragg bandgaps is studied and the possibility of creating a single combined bandgap is considered.

**METHOD**

The conventional sandwich construction involves bonding two thin facesheets on either side of a thicker, lightweight core material. Typically, facesheets provide the bending rigidity while the shear stiffness is provided by the core.

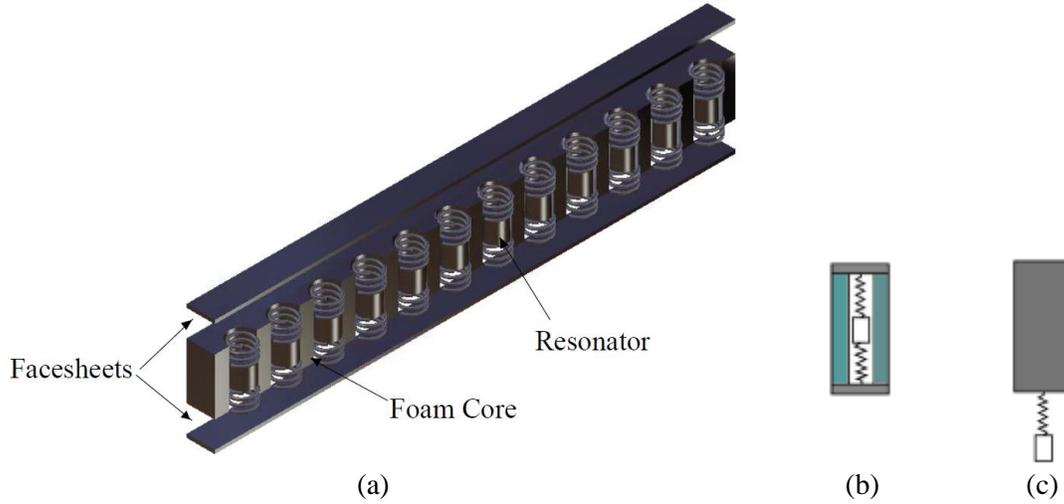

(a)                (b)      (c)

**Figure 1:** (a) Schematic of a sandwich beam with internal resonators; (b) unit cell; (c) equivalent unit cell modeled as a Timoshenko beam with attached resonators.

Consider a sandwich beam of width b and cross-sectional moment of inertia I made using facesheets of thickness $h_f$, elastic modulus $E_f$, and density $\rho_f$, and core of thickness $h_c$, shear modulus $G_c$, and density $\rho_c$. For long wavelengths it can be safely assumed that the facesheets are solely responsible for the bending behavior while the shear behavior depends completely on the core shear properties [21]. Such a sandwich beam can be represented as a Timoshenko beam with its equivalent bending rigidity EI, shear rigidity GA, mass per unit length $\rho A$, and rotary inertia $\rho I$ calculated as [21]

$$EI = E_f b(h_c^2 h_f/2 + h_c h_f^2) \tag{1}$$

$$GA = G_c b(h_c + 2h_f) \tag{2}$$

$$\rho A = 2\rho_f b h_f + \rho_c b h_c \tag{3}$$

$$\rho I = \rho_f b(h_c^2 h_f/2 + h_c h_f^2) + \rho_c b h_c^3/12 \tag{4}$$

Using the Timoshenko beam theory, the equations of motion for this beam are

$$GA[v''(x,t) + \varphi'(x,t)] - \rho A\ddot{v}(x,t) = 0 \qquad (5)$$

$$EI\varphi''(x,t) - GA[v'(x,t) + \varphi(x,t)] - \rho A\ddot{\varphi}(x,t) = 0 \qquad (6)$$

where $v$ is the transverse displacement of the sandwich beam and $\varphi$ is the cross-sectional rotation.

Assume harmonic solutions of the form:

$$v = v_o e^{-i(kx - \omega t)}, \qquad\qquad \varphi = \varphi_o e^{-i(kx - \omega t)} \qquad (10)$$

Then, the characteristic equation can be derived to be [27]:

$$k^4 - \left[\frac{\rho I}{EI} + \frac{\rho A}{GA}\right]\omega^2 k^2 + \left[\frac{\rho I \rho A}{EIGA} - \frac{\rho A}{EI}\right]\omega^2 = 0 \qquad (11)$$

In the frequency domain, the general solution may be written as:

$$\hat{v}(x) = Ae^{-ik_1 x} + Be^{-ik_2 x} + Ce^{ik_1 x} + De^{ik_2 x} \qquad (12)$$

$$\hat{\varphi}(x) = \hat{A}e^{-ik_1 x} + \hat{B}e^{-ik_2 x} + \hat{C}e^{ik_1 x} + \hat{D}e^{ik_2 x}$$

For an infinitely long Timoshenko beam excited by a point harmonic force at x = 0, using symmetry and antisymmetry conditions, we write the displacements and rotations to the right and left-hand parts of the beam as

$$\widehat{v_R}(x) = B_1 e^{-ik_1 x} + B_2 e^{-ik_2 x} \ \ (x \geq 0), \widehat{v_L}(x) = B_1 e^{ik_1 x} + B_2 e^{ik_2 x} \ \ (x \leq 0) \qquad (13)$$

$$\widehat{\varphi_R}(x) = C_1 e^{-ik_1 x} + C_2 e^{-ik_2 x} \ \ (x \geq 0), \ \ \widehat{\varphi_L}(x) = C_1 e^{ik_1 x} + C_2 e^{ik_2 x} \ \ (x \leq 0) \qquad (14)$$

where the amplitude coefficients are related as

$$C_j = i\left[\frac{GAk_j^2 - \rho A\omega^2}{GAk_j}\right]B_j, \qquad \text{for } j = 1,2 \qquad (15)$$

The boundary conditions to be satisfied are as follows: (a) due to antisymmetry of φ about x = 0; the rotations $\varphi_R(0) = \varphi_L(0) = 0$; and (b) the shear force, V, on the beam cross-section just to the right or left of x = 0 is equal to half the applied force. The application of the first boundary condition leads to $C_1 = - C_2$. To apply the other boundary condition we use the expression for the shear force on the beam cross-section:

$$V = -GA\left[\frac{\partial v}{\partial x} - \varphi\right] \tag{16}$$

These boundary conditions lead to

$$B_1 = \frac{P}{2iGA}\frac{k_1}{k_s^2}\left[\frac{k_2^2 - k_s^2}{k_2^2 - k_1^2}\right], \qquad B_2 = \frac{-P}{2iGA}\frac{k_2}{k_s^2}\left[\frac{k_1^2 - k_s^2}{k_2^2 - k_1^2}\right] \tag{17}$$

where $k_s = \omega\sqrt{\frac{\rho A}{GA}}$, and P is the applied load.

Thus, the transverse displacement at any point is:

$$\hat{v}(x) = \frac{P}{2iGA}\left[\frac{k_1}{k_s^2}\left[\frac{k_2^2 - k_s^2}{k_2^2 - k_1^2}\right]e^{-ik_1 x} - \frac{k_2}{k_s^2}\left[\frac{k_1^2 - k_s^2}{k_2^2 - k_1^2}\right]e^{-ik_2 x}\right] \tag{18}$$

An infinite beam with periodically placed resonators can be viewed as a uniform structure on which the resonators impose forces at regular intervals. The force applied by one attached spring-mass system on the infinite beam is easily calculated and can be conveniently used to compute and study the wave motion in the whole periodic structure under free or forced harmonic conditions. Since the response of a point in a periodic structure is related to a corresponding point in the adjacent section through the propagation constant [4], it follows that the forces and moment exerted by the periodic constraints are similarly related to each other through the propagation constant. Thus, they can be considered to be an array of forces phased through e$^{-i\mu}$, where μ is the propagation constant [4]. In general, μ is complex and written as

$$\mu = \delta + i\gamma \tag{19}$$

where the real part of μ (i.e. δ) carries information about the phase difference between the responses measured at two points one periodic length apart (L), while the imaginary part (i.e. γ) describes the rate of decay of the amplitude between these two points. The wavenumber of a propagating wave (k), i.e. for a wave with γ = 0, is given as

$$k = \delta/\text{L} \tag{20}$$

and the corresponding wavelength λ is

$$\lambda = 2\pi/\text{k} \ = \ 2\pi\text{L}/\delta \tag{21}$$

Consider the above beam to be subjected to a phased array of harmonic forces, P$_j$ = P$_0$e$^{-ij\mu}$e$^{-i\omega t}$, applied with periodicity L. Assuming only the force P$_0$ to be active, the transverse displacement to the right, v$_r$(x), and to the left, v$_l$(x), of the force can be expressed as:

$$v_r(x) = P_0 \sum_{j=1}^{2}\left[a_j e^{-ik_j x}\right], \ \ v_l(x) = P_0 \sum_{j=1}^{2}\left[a_j e^{ik_j x}\right] \tag{22}$$

where k$_1$ and k$_2$ are solutions of the characteristic equation (Eq. 11), and using Eq. 18, a$_1$ and a$_2$ are written as:

$$a_1 = \frac{1}{2iGA}\frac{k_1}{k_s^2}\left[\frac{k_2^2 - k_s^2}{k_2^2 - k_1^2}\right], \qquad a_2 = \frac{-1}{2iGA}\frac{k_2}{k_s^2}\left[\frac{k_1^2 - k_s^2}{k_2^2 - k_1^2}\right] \tag{23}$$

The transverse displacement at the location x=0, due to the force acting at that point is

$$v(0) = P_0[a_1 + a_2] \tag{24}$$

and the displacement at a distance rL to the right or left of the force is

$$v(rL) = P_0 \sum_{j=1}^{2} \left[ a_j e^{-ik_j rL} \right] \tag{25}$$

Taking all the forces into consideration, the total response at x = 0 is the sum of all the wave fields generated by the infinite phased array, i.e., the response due to $P_0$ + the response due to all the forces to the right of $P_0$, and the response due to all the forces to the left of $P_0$.

Thus we have

$$v_{op} = P_0 \sum_{j=1}^{2} [a_j] + \sum_{r=1}^{\infty} P_r \sum_{j=1}^{2} \left[ a_j e^{-ik_j rL} \right] + \sum_{s=1}^{\infty} P_s \sum_{j=1}^{2} \left[ a_j e^{-ik_j sL} \right] \tag{26}$$

Using the relations $P_r = P_0 e^{-i\mu r}$ and $P_s = P_0 e^{i\mu s}$ we write

$$v_{op} = P_0 \sum_{j=1}^{2} [a_j] + P_0 \sum_{j=1}^{2} a_j \sum_{r=1}^{\infty} \left[ e^{-ik_j rL + i\mu r} + e^{-ik_j rL - i\mu r} \right] \tag{27}$$

This expression is further simplified using the identity $\sum_{r=1}^{\infty} e^{cr} = \frac{e^c}{1 - e^c}$ and by using Euler's formula to replace the complex exponentials with trigonometric functions as

$$v_{op} = P_0 \sum_{j=1}^{2} a_j \left\{ \frac{i \sin k_j L}{\cos k_j L - \cos \mu} \right\} \tag{28}$$

For resonators with stiffness K and mass m, the transverse forces due to the resonators are given as [25] $P_r = -K_T v_r, \quad P_0 = -K_T v_0$, etc. where

$$K_T = K - \left( \frac{K^2}{K - m\omega^2} \right) \tag{29}$$

Thus, the effective force at x = 0 due to the array of resonators is obtained as the sum:

$$P_0 = K_T P_0 \sum_{j=1}^{2} a_j \left\{ \frac{i \sin k_j L}{\cos k_j L \ - \cos \mu} \right\} \tag{30}$$

Plugging in the values of $a_1$ and $a_2$ derived above and solving for cosμ, we obtain the quadratic equation:

$$\cos^2 \mu \ - \beta_1 \cos \mu + \beta_2 = 0 \tag{31}$$

where,

$$\beta_1 = iK_T[a_1 \sin k_1 L + a_2 \sin k_2 L] - [\cos k_1 L + \cos k_2 L] \tag{32}$$

$$\beta_2 = iK_T[a_1 \sin k_1 L \cos k_2 L + a_2 \sin k_2 L \cos k_1 L] + [\cos k_1 L \cos k_2 L] \tag{33}$$

The solution to this quadratic equation gives the propagation constant, thus allowing us to analyze the beam's wave propagation behavior.

Dispersion curves obtained using the phased array approach are compared with those obtained using the volume averaging method and finite element method. Details about the volume averaging technique can be obtained in [21]. The finite element method is based on the observation that harmonic wave propagation can be effectively viewed as free harmonic vibration with sinusoidal mode shapes [27]. For a sinusoidal propagating wave, the wave nodes are effectively described as being simply supported. Using this fact, and taking advantage of the Timoshenko beam assumption, 2-node linear Timoshenko (shear flexible) beam elements (B21) with a meshed cross-section are used to model the sandwich beam using Abaqus finite element software. The resonator masses are modeled as point-masses, m, and are attached to the beam nodes by using linear spring elements with appropriate spring stiffness, K, to obtain the chosen local resonance frequency which is given as, $\omega_r = \sqrt{K/m}$. Using an eigenvalue analysis, the natural frequencies, $\omega_n$, and the characteristic wavelengths, $\lambda_n$, are obtained. The obtained wavelengths are then used to calculate the wavenumber (k = 2π/λ) and the dispersion curve is subsequently plotted. Note that only the real part of the wavenumber can be obtained using such a finite element model, which is used to visualize the bandgap region. Attenuation of harmonic waves in the predicted wave attenuation bandgaps is further verified using transient simulations performed using Abaqus/Explicit. A 100 m long beam, with the length chosen so as to avoid end reflection interference effects, is subjected to a single frequency harmonic displacement at one end while the other end is kept free. To achieve accuracy and ensure convergence a double precision analysis is employed and each unit cell is modeled using ten Timoshenko beam elements. The output displacement histories for different loading frequencies are subsequently obtained and compared with the input displacement to assess the attenuation of the waves.

## RESULTS AND DISCUSSION

### *Verification and comparison*

The dispersion curves obtained using the phased array and the volume averaging methods are compared with those obtained using finite element simulations in Figure 2, where $\Omega$ is the frequency normalized with respect to the local resonance frequency, while kL is the wavenumber normalized with respect to the unit cell length. To maintain validity of the equivalent Timoshenko beam theory assumption, the analysis was restricted to low frequencies and local resonance frequency of 200 Hz was chosen. The beam material and geometrical properties, and the resonator mass and stiffness are summarized in Table 1(a) and (b). Note that the same beam material and geometrical properties are used to obtain all the results presented in this paper.

**Table 1(a):** Sandwich beam material and geometrical properties

| $E_f$ (GPa) | $h_f$ (mm) | $\rho_f$ (Kg/m$^3$) | $\rho_c$ (Kg/m$^3$) | $h_c$ (mm) | $G_c$ (MPa) | b (mm) |
|---|---|---|---|---|---|---|
| 85.42 | 1.2 | 1570 | 96.11 | 25 | 1 | 25 |

**Table 1(b):** Equivalent Timoshenko beam and resonator properties

| EI (Pa*m$^4$) | GA (Pa*m$^2$) | $\rho A$ (Kg/m) | $\rho I$ (Kg*m) | K (N/m) | m (Kg) |
|---|---|---|---|---|---|
| 877.6905 | 625 | 0.1543 | $1.926 \times 10^{-5}$ | $7.1308 \times 10^{5}$ | 0.025 |

Figure 2(a) compares the dispersion curves obtained for beam with resonators placed periodically at a distance of 25 mm, i.e. L = 25 mm. All three solutions predict a wave attenuation bandgap in the frequency region near the local resonance frequency. This bandgap arises due to the interaction between the incident waves and the resonators and is termed as a local resonance (LR) bandgap [21]. The curves obtained using the phased array method match the finite element solution exactly while the volume averaging method solution diverges from the other two near the bandgap bounding frequencies. The error is further increased when L is increased to 50 mm, as shown in Figure 2(b).

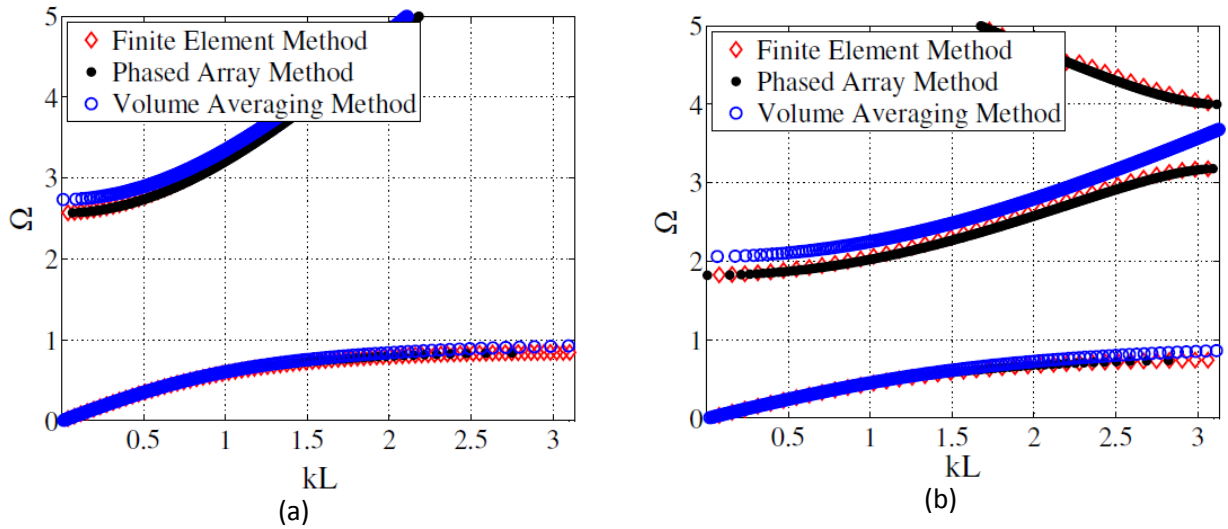

**Figure 2:** Dispersion curves for a sandwich beam with resonators inserted with periodicity (a) L = 25 mm and (b) L = 50 mm.

As the resonator spacing is increased, the finite element and phased array method curves predict the existence of a higher order bandgap which is absent in the dispersion curve obtained using the volume averaging method. This wave attenuation bandgap is associated with the periodicity of the structure and is due to the interference effects introduced due to the periodicity of the system and is termed as a Bragg bandgap [4]. The volume averaging method assumes a uniform distribution of resonators and uses a continuous field variable to allow analysis of the system using Hamilton's principle. It ignores the effect of the resonator periodicity on the wave propagation behavior of the system and is unable to predict the existence of Bragg bandgaps. Thus, the volume averaging technique is suitable only for systems with extremely close-spaced resonators which effectively moves the Bragg bandgaps outside the frequency range of interest and thus safely ignores the effect of resonator periodicity.

The presence of the predicted bandgaps is verified using transient simulations performed using Abaqus/Explicit. The displacement histories for different loading frequencies obtained 2 m away from the input end are plotted in Figure 3.

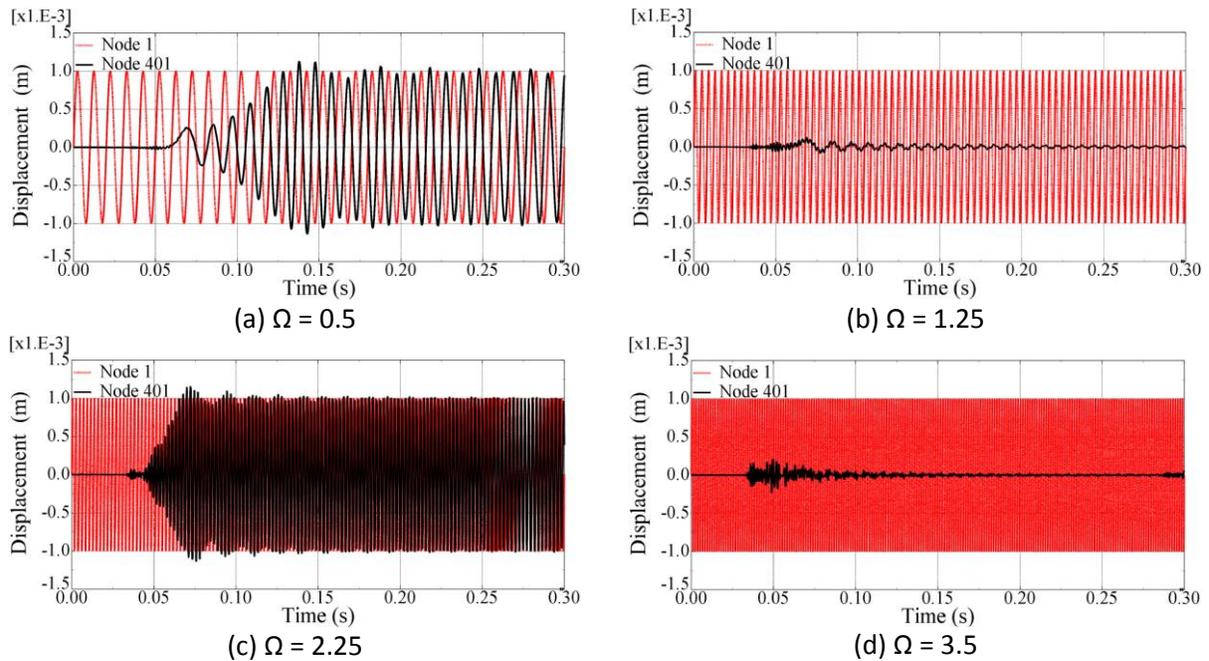

**Figure 3:** Displacement histories at the input (Node 1) and 2 m away (Node 401) for a sandwich beam subjected to single frequency harmonic displacements.

Four different input displacement frequencies are chosen to demonstrate the wave behavior in the four different zones shown in Figure 2(b). For $\Omega = 0.5$ and 2.25 (Figures 3(a) and 3(c)), which lie in the passing band, no wave attenuation is obtained and after an initial transient part the displacement output shows a steady state magnitude equal to the input. The wave behavior for $\Omega = 1.25$ and 3.5 (Figures 3(b) and 3(d)), both of which lie in the wave attenuation zones as predicted by the phased array and finite dispersion curves, shows the expected attenuation and the output displacement magnitude is extremely diminished as compared to the input displacement. It should be noted that the volume averaging method is unable to predict the attenuation at $\Omega = 3.5$ and incorrectly identifies it as a region inside a passing band.

*Representation of bandgap bounding frequencies*

Often, when analyzing the wave propagation behavior of a structure, the wave attenuation start and stop frequencies are of the primary concern. Mead [28] demonstrated that these frequencies, commonly referred to as bounding frequencies, of a periodic system are equal to the modal frequencies of the system unit cell with appropriate boundary conditions. For an infinite sandwich beam with periodically inserted resonators two different unit cells are viable; with resonator located at the center of the unit cell or with resonators at the unit cell boundaries with the mass and spring stiffness equally divided. Both configurations are shown in Figure 4.

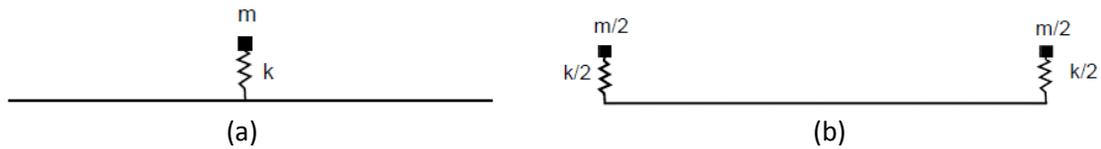

**Figure 4:** Possible unit cell configurations.

The relationship between the modal frequencies of the unit cell type and its associated boundaries, with the bounding frequencies obtained in the dispersion curves are studied using finite element models. Two boundary conditions are found to be relevant; simply supported and sliding boundaries. All the bounding frequencies are found to match the modal frequencies of one of the following three configurations: (a) simply supported unit cell with resonators at the ends, (b) sliding - sliding unit cell with resonator at the center and (c) simply supported unit cell with resonator at the center. The relationships are summarized in Table 2, where for convenience the mode shapes of the unit cell are also shown.

**Table 2:** Relationship between bounding frequencies and unit cell mode shapes

| | | | |
|---|---|---|---|
| 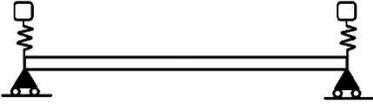 | 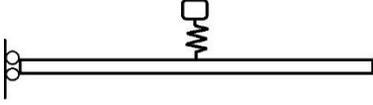 | | 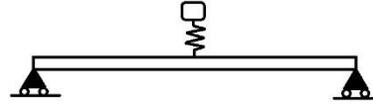 |
| **(a)** | **(b)** | | **(c)** |

| Bandgap # | Bounding frequency type | Configuration type | Mode shape |
|---|---|---|---|
| First (LR) | Cut-on (147.84 Hz) | (c) | 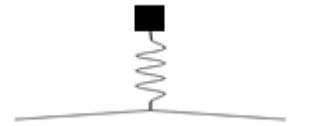 |
| | Cut-off (364.36 Hz) | (b) | |
| Second (Bragg) | Cut-on (635.68 Hz) | (a) | 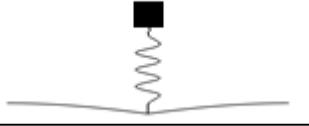 |
| | Cut-off (800.42 Hz) | (c) | |
| Third (Bragg) | Cut-on (1270.82 Hz) | (a) | 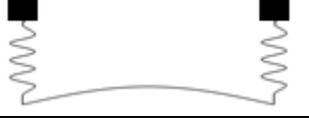 |
| | Cut-off (1368.38 Hz) | (b) | |

Both bounding frequencies for the LR bandgap correspond to modal frequencies of a unit cell with the resonator attached at the center, with the cut-on and cut-off frequencies corresponding to the simply supported and sliding condition, respectively. Thus, both bounding frequencies for this bandgap are dependent not only on the resonator parameters (m and K) but also on the spacing (L) between the resonators. For the Bragg bandgap, the cut-on frequencies always correspond to the modal frequency of a simply supported unit cell with the resonators at the two ends. Such a configuration is equivalent to a beam unit cell without any resonators since the resonator-beam attachment points act as nodes and the resonators remain stationary. Thus, as reported for other similar systems [28], the cut-on frequency of the Bragg bandgap is dependent solely on the periodicity of the system (L) and is independent of the resonator parameters. Conversely, the Bragg bandgap cut-off frequencies always correspond to a unit cell with resonators attached at the center with the end conditions alternating between simply supported and sliding configurations. Hence, the Bragg bandgap cut-off frequencies are influenced by the system periodicity as well as the resonator parameters. This allows the possibility of tailoring not only the LR bandgap, but also the width of the Bragg gaps introduced into the system due to the resonator periodicity.

Figures 5(a)-(d) show the influence of resonator mass and stiffness on the real and imaginary part of the propagation constant for a beam with resonators tuned to a specific frequency. For a fixed local resonance frequency, both the bandgap widths increase as either the resonator mass or stiffness are increased. As per the discussion above, both bounding frequencies for the local resonance bandgap are affected by a change in resonator parameters. Increasing the resonator mass or stiffness causes the cut-on frequency to slowly decrease while the cut-off frequency increases much more rapidly as the bandgap retains an asymmetric structure [15] around the frequency of maximum attenuation, i.e. the local resonance frequency. For the Bragg bandgap, the cut-on frequency remains unaffected while the cut-off frequency increases gradually causing the bandgap to widen. An increase in resonator mass or stiffness also causes an increase in the magnitude of attenuation in the Bragg bandgap while it retains its symmetric structure around a central maximum attenuation value. This demonstrates the tunability of the Bragg bandgap cut-off frequency by varying the resonator parameters alone, with the resonator mass being more effective in changing the bounding frequencies as compared to the stiffness, as can be noticed by observing the scaling of the mass and stiffness axes.

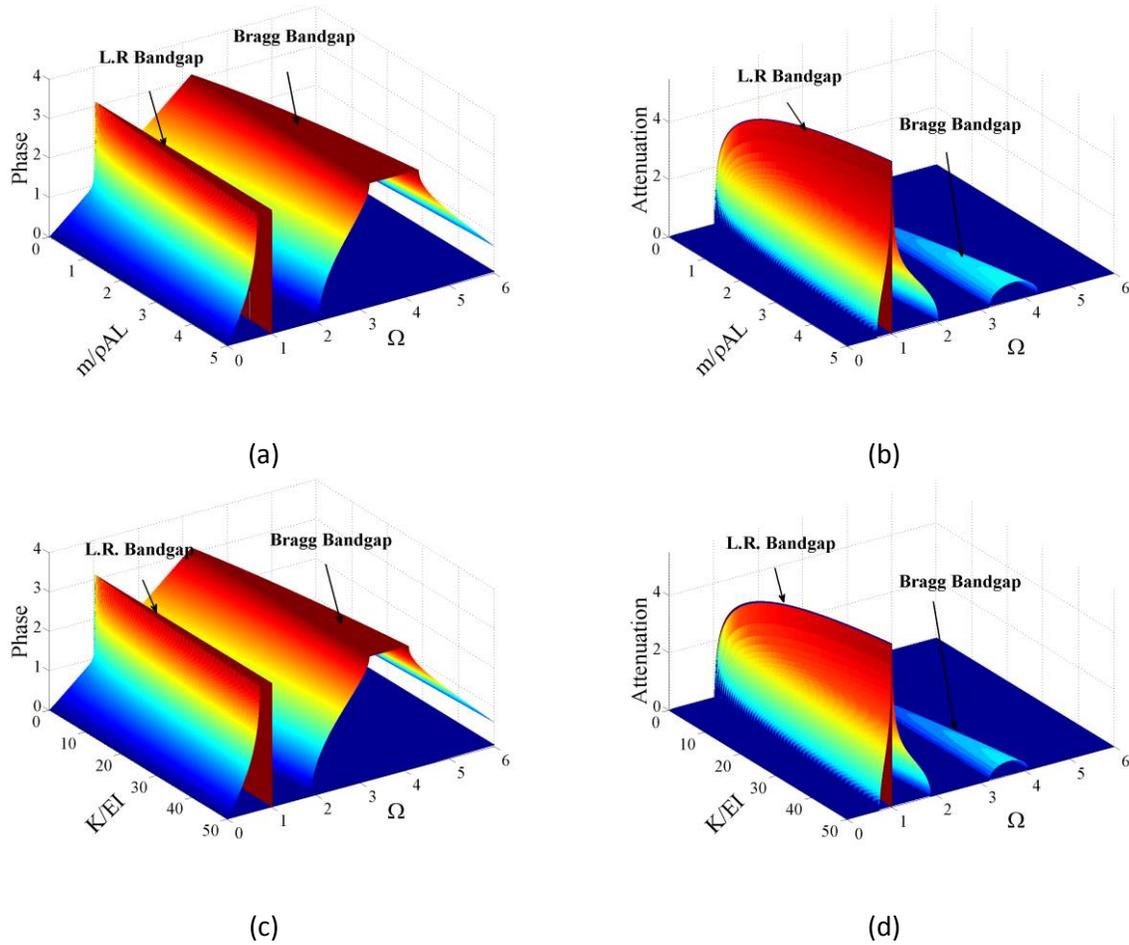

(a)                                        (b)

(c)                                        (d)

**Figure 5 (a - d)**: Effect of resonator mass and stiffness on the bandgap behavior with the local resonance frequency kept fixed (200 Hz). Figures 5(a) and 5(b) show the effect of mass variation on the phase (real) and attenuation (imaginary) part of the phase constant, respectively, while Figures 5(c) and 5(d) show the effect of stiffness variation on the phase (real) and attenuation (imaginary) part of the phase constant, respectively. Note that the mass axis is normalized with respect to the beam unit cell mass, while the stiffness axis is normalized with respect to the beam bending stiffness.

Modal frequencies of unit cells shown in Table 2 decrease as the unit cell length is increased. Figure 6(a)-(d) shows the influence of resonator spacing on the bandgap structure. Bounding frequencies for both the LR as well as the Bragg bandgap decrease with increasing resonator spacing. Due to the nature of the associated boundary conditions, the cut-on frequency of the local resonator decreases much more gradually than the cut-off frequency and the width of the bandgap noticeably decreases with increasing L. For the symmetrical Bragg band gap, as L increases both the bounding frequencies decrease with a gradual decrease in bandgap width, while the magnitude of maximum attenuation increases. The gap shifts to a lower frequency range with increasing L and multiple Bragg band gaps can be seen to emerge within the frequency range under consideration. Thus, by increasing the resonator spacing multiple wave attenuation band gaps can be obtained at lower frequencies.

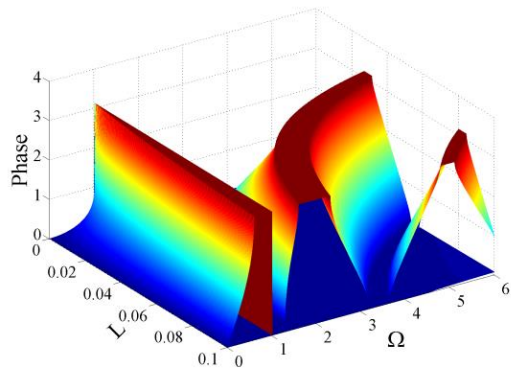

(a)

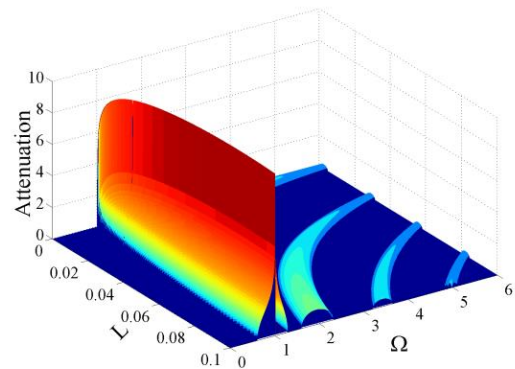

(b)

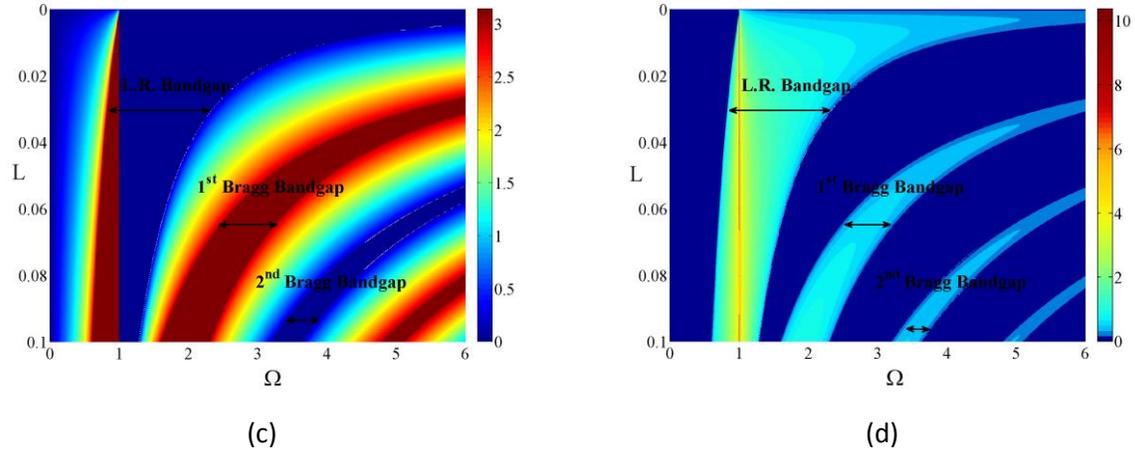

**Figure 6 (a - d)**: Effect of resonator periodicity on the bandgap behavior for a constant local resonance frequency (200 Hz). Figures 6(a) and 6(b) show the effect of periodicity on the phase (real) and attenuation (imaginary) part of the phase constant respectively. Figures 6(c) and 6(d) show the planar view of Figures 6(a) and 6(b), with the LR and Bragg bandgaps marked out.

*Bandgap interaction*

It is clear from the above analysis that though the LR and Bragg bandgaps are created due to two different physical mechanisms, they are still both dependent on the resonator parameters and the unit cell length. Here, we study the interaction between the two bandgaps and the possibility of creation of a single 'Resonant-Bragg' [12] bandgap.

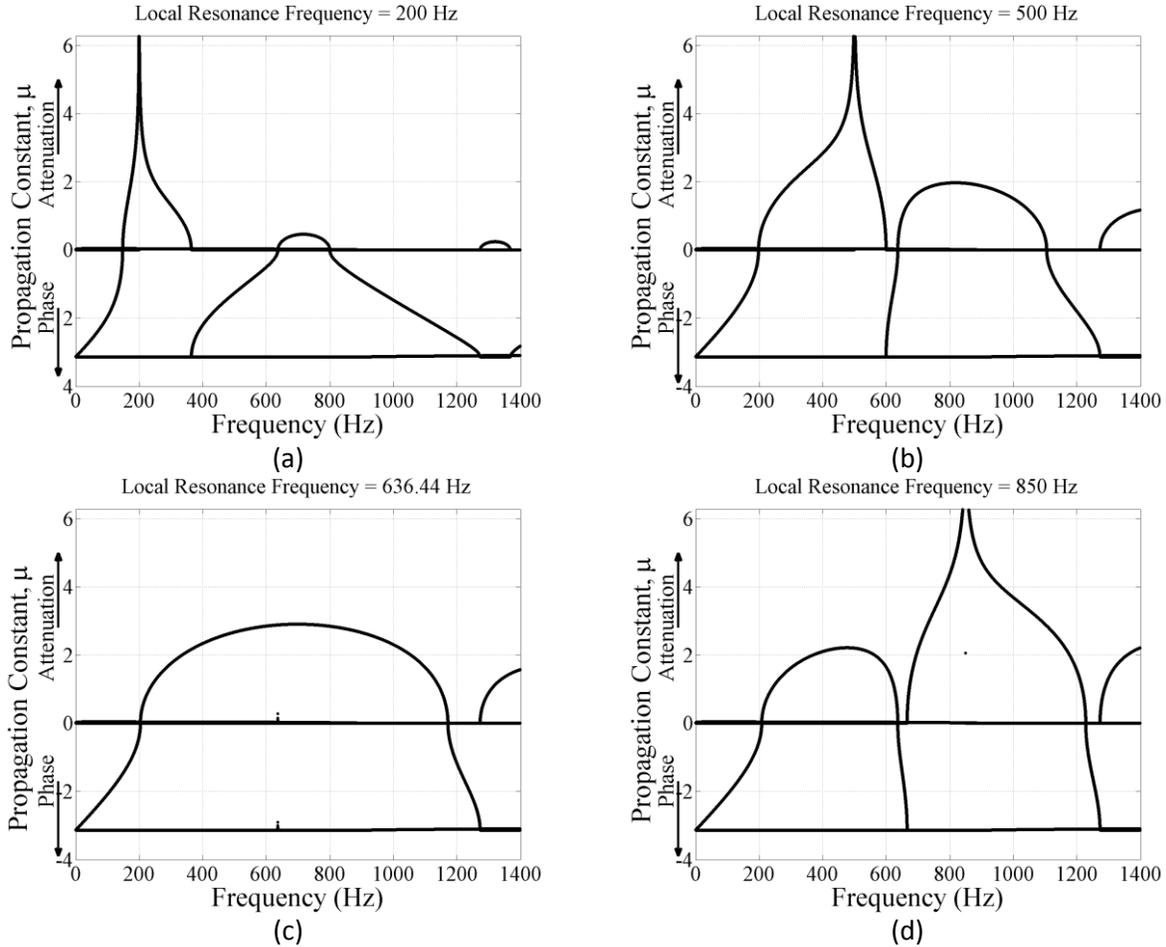

**Figure 7:** Propagation constants for beam with resonators tuned at **(a)** 200 Hz, **(b)** 500 Hz, **(c)** 636.44 Hz, and **(d)** 850 Hz, respectively. The periodicity is kept constant at L = 0.05 m. For all cases the resonator mass is kept constant at 25 gm and the local resonance frequency is varied by varying the resonator stiffness accordingly.

The interaction between the two bandgaps and their influence on each other is studied by varying the LR bandgap position. This is achieved by varying the resonator mass while keeping L fixed and thus shifting the position of the local resonance bandgap relative to the Bragg bandgap cut-on frequency. For bandgaps located sufficiently apart (Figure 7(a)), no bandgap interaction is noticed. As the local resonance bandgap is moved closer to the Bragg bandgap (Figure 7(b)), the structures of the attenuation constants for the two bandgaps noticeably change. For the Bragg bandgap, the attenuation constant is no longer symmetric around the mid-gap frequency and attains a maximum attenuation value at a lower frequency. Also, the nature of asymmetry associated with the LR bandgap changes and it now spreads out more towards frequencies below the local resonance frequency. Figure 7(c) shows the band structure obtained when the local resonance frequency matches the first Bragg bandgap cut-on frequency. Under this condition, researchers investigating other systems, [9], [12], [13], and [14], have reported merging of the two bandgaps causing a very wide and strongly attenuating single hybrid 'Resonance-Bragg' bandgap. However, in our studies, such a single bandgap was not found to exist. Though the two bandgaps seem to couple together and create an almost-symmetric, wide bandgap, a narrow passing band separating the two gaps still exists. As the local resonance frequency is increased further and located inside the original Bragg bandgap, the band structure noticeably changes. The Bragg bandgap 'flips' around the original cut-on frequency, which is the new Bragg bandgap cut-off frequency, and cut-on at a much lower frequency. As the local resonance frequency is increased further, the Bragg cut-off frequency remains unaffected while the cut-on frequency gradually increases. Thus, the new cut-off frequency is independent of resonator mass and stiffness and hence equivalent to the cut-on frequency for the Bragg bandgap obtained when the local resonance frequency is below the classical Bragg frequency. This Bragg cut-on to cut-off frequency switching can be seen more clearly in Figures 8(a)-(c) which focus on the behavior of the attenuation constant in the proximity of the first Bragg frequency.

This change in behavior of the Bragg bandgap is verified by looking at the unit cell modal frequencies associated with the new bounding frequencies. Table 3 summarizes the mode shapes and frequencies associated with the bounding frequencies for the system analyzed in Figure 7(d).

As predicted, the cut-on frequency for the Bragg bandgap is a resonator dependent mode and it matches the cut-on frequency mode earlier associated with the LR bandgap cut-on frequency.

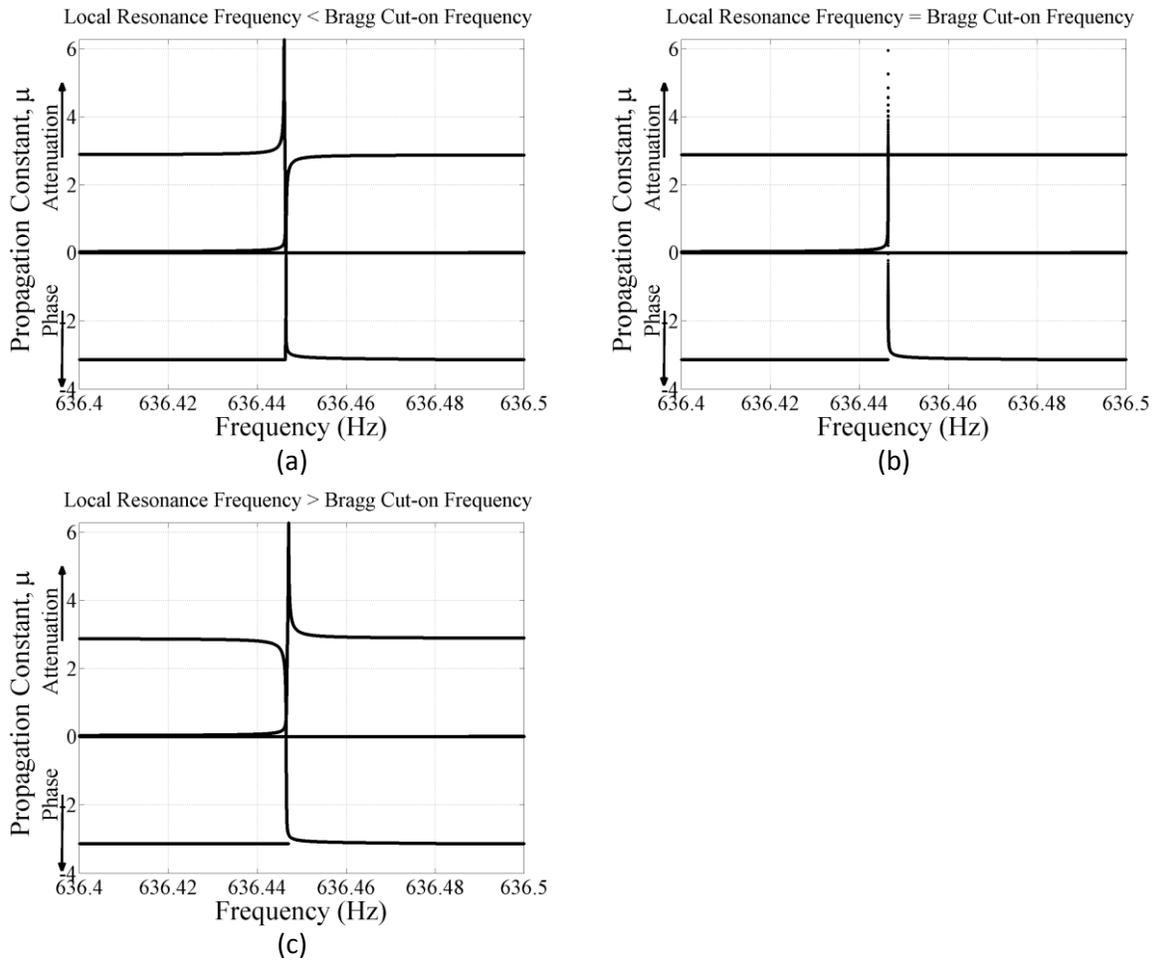

**Figure 8 (a – c):** Bandgap transition as the local resonance approaches the Bragg bandgap cut-on frequency. When the local resonance frequency matches the Bragg frequency (Figure 8(b)) the two bandgaps appear to merge, though a narrow passband still persists.

Also, the cut-off frequency mode is the mode earlier associated with the Bragg cut-on frequency and is thus independent of the resonator mass and stiffness and dependent only on the unit cell length. Both the bounding frequency modes for the LR bandgap are still resonator dependent, with the earlier cut-off mode being the new cut-on mode and the earlier first Bragg bandgap cut-off mode being the new LR bandgap cut-off mode. The complete interaction behavior of the bandgaps is summarized in Figures 9(a)-(d).

**Table 3:** Relationship between bounding frequencies and unit cell mode shapes when the local resonance frequency is greater than the first Bragg frequency.

| Bandgap # | Bounding frequency type | Configuration type | Mode shape |
|---|---|---|---|
| | 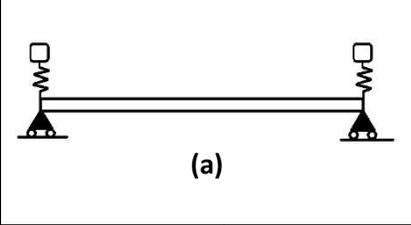 (a) | 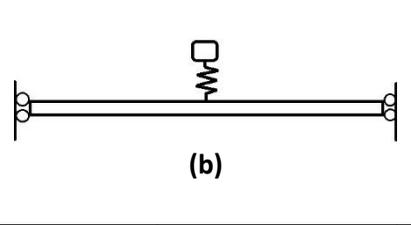 (b) | 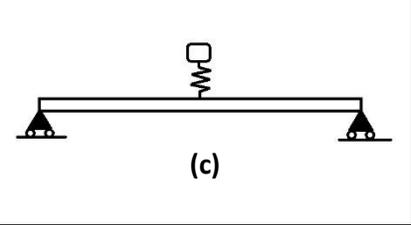 (c) |
| First (Bragg) | Cut-on (208.2 Hz) | (c) | 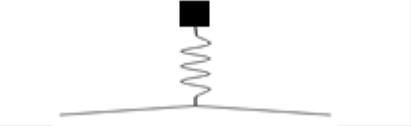 |
| | Cut-off (636.4 Hz) | (b) | 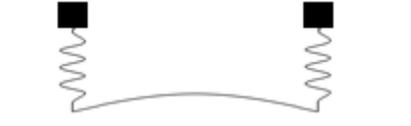 |
| Second (LR) | Cut-on (665.8 Hz) | (a) | 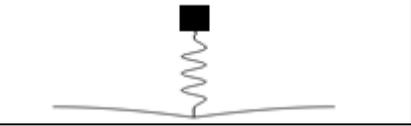 |
| | Cut-off (1228 Hz) | (c) | 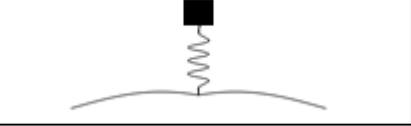 |

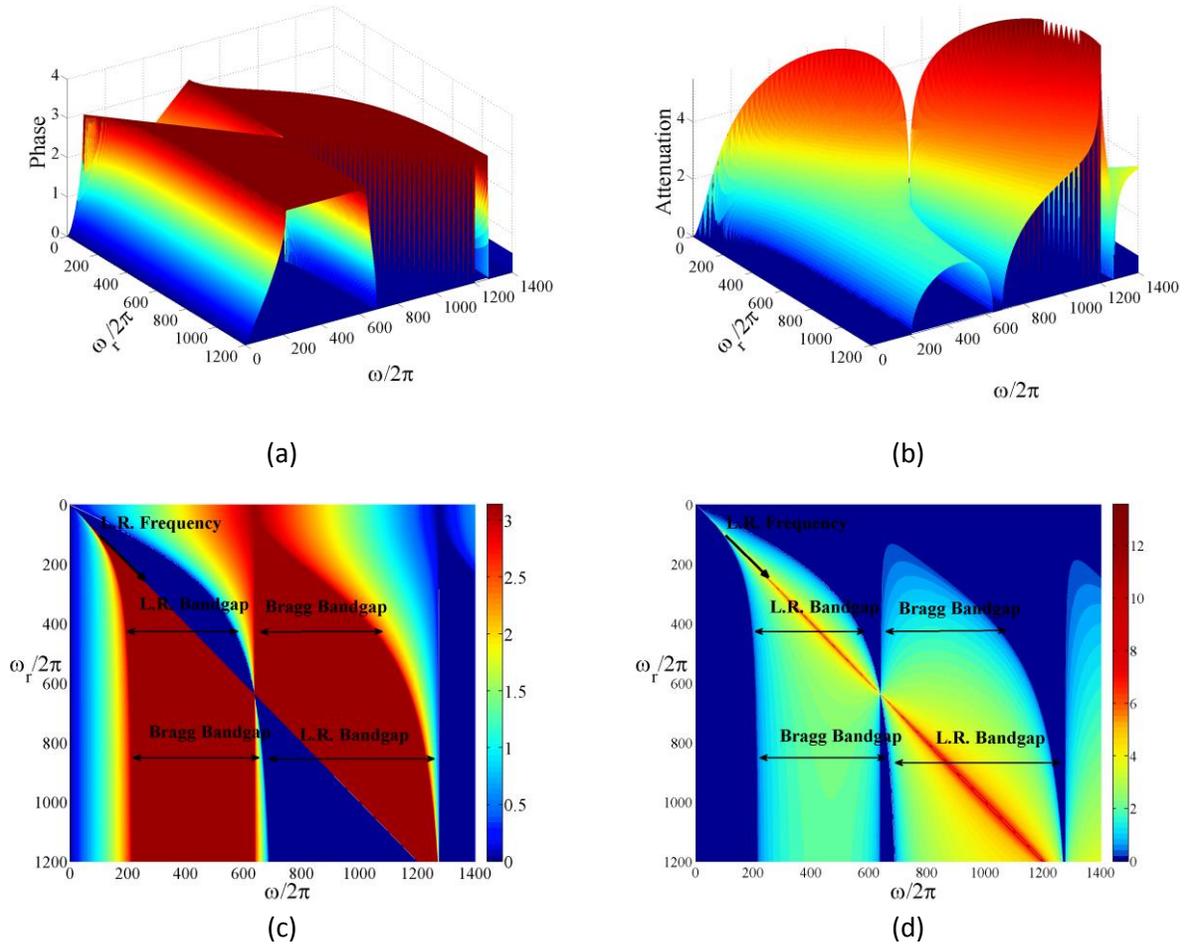

**Figure 9 (a - d)**: Effect of local resonance frequency on the bandgap behavior with a constant periodicity (L = 0.05 m). Figures 9(a) and 9(b) show the effect of local resonance variation on the phase (real) and attenuation (imaginary) part of the phase constant respectively. Figures 9(c) and 9(d) show the planar view of Figures 9(a) and 9(b), with the bandgaps marked out.

## CONCLUSION

The wave attenuation behavior of an infinite sandwich beam with periodically inserted resonators has been studied. A closed form solution for the propagation constant for such a beam was obtained under the long wavelength assumption by using a phased-array approach. The obtained solution shows the presence of attenuation bandgaps in addition to the bandgap induced due to the local resonance of the inserted resonators. These Bragg bandgaps are induced due to resonator periodicity and are not captured when analyzing such systems using a volume averaging technique.

The dependence of the bounding frequencies of these bandgaps on resonator mass, stiffness and periodicity was investigated using simple unit cell models with appropriate boundary conditions. Though the LR bandgap and the Bragg bandgap are induced due to two distinct physical phenomena, the bounding frequencies of both the bandgaps are dependent on the inserted resonators. For the LR bandgap, both bounding frequencies as well as the magnitude of attenuation are dependent on the resonator mass, stiffness as well as their periodicity. For the Bragg bandgap, though the cut-off frequencies and magnitude of attenuation are dependent on the resonator mass, stiffness and periodicity, the cut-on frequencies are dependent only on the periodicity and are independent of the resonator mass and stiffness. An increase in resonator mass or stiffness causes both bandgaps to widen, whereas an increase in the distance between successive resonators causes the bandgaps to shrink.

The interaction between the two bandgaps and the possibility of creation of a single combined high attenuation bandgap was considered. When the local resonance frequency coincides with the Bragg cut-on frequency, a pseudo-combined gap with a narrow pass band was obtained. It was found that the two bandgaps cannot be combined into a single bandgap and that a passing band exists between the two gaps under all conditions. The width of this passing band is minimum when the local resonance frequency coincides with the Bragg cut-on frequency. When the local resonance frequency is greater than a Bragg bandgap cut-on frequency, the Bragg bandgaps were found to transition into sub-wavelength bandgaps which cut-off instead of cutting-on at frequencies given by the classical Bragg frequencies. Thus, the periodicity of the inserted resonators plays a significant role in the resultant wave attenuation behavior and the location and width of Bragg bandgaps may also be tailored by choosing appropriate resonator properties and periodicity.

**FUNDING**

This work was supported by the Office of Naval Research [grant number N00014-11-1-0580]. Dr. Yapa D.S. Rajapakse was the program manager.

**REFERENCES**

[1] Z. Liu, X. Zhang, Y. Mao, Y. Y. Zhu, Z. Yang, C. T. Chan, P. Sheng, Locally resonant sonic materials. Science, 289 (5485) (2000), 1734-1736. doi:10.1126/science.289.5485.1734.

[2] M. I. Hussein, M. J. Leamy, M. Ruzzene, Dynamics of phononic materials and structures: Historical origins, recent progress and future outlook. Applied Mechanics Reviews, 66 (2014), 040802. doi:10.1115/1.40262911.

[3] H. H. Huang, C. T. Sun, Wave attenuation mechanism in an acoustic metamaterial with negative effective mass density. New Journal of Physics, 11 (2009), 013003. doi:10.1088/1367-2630/11/1/013003.

[4] L. Brillouin, Wave propagation in periodic structures, Dover Publications, New York, 1953.

[5] Lord Rayleigh, On the maintenance of vibrations by forces of double frequency, and on the propagation of waves through a medium endowed with a periodic structure. Philosophical Magazine XXIV (1887), 145-159.

[6] D. J. Mead, Wave propagation in continuous periodic structures: Research contributions from Southampton, 1964 – 1995. Journal of Sound and Vibration, 190 (3) (1996), 495-524. doi:10.1006/jsvi.1996.0076.

[7] M. Rupin, F. Lemoult, G. Lerosey, P. Roux, Experimental demonstration of ordered and disordered multiresonant metamaterials for lamb waves, Physics Review Letters, 112 (2014), 234301. doi:10.1103/PhysRevLett.112.234301.

[8] T. Still, W .Cheng, M. Retsch, R. Sainidou, J. Wang, U. Jonas, N. Stefanou, G. Fytas, Simultaneous occurrence of structure-directed and particle-resonance-induced phononic gaps in colloidal films. Physical Review Letters, 100 (2008), 194301. doi:10.1103/PhysRevLett.100.194301.

[9] C. Croënne, E. J. S. Lee, H. Hu, J. H. Page, Band gaps in phononic crystals: Generation mechanisms and interaction effects. AIP Advances 1 (2001), 041401. doi:10.1063/1.3675797.

[10] Y. Achaoui, A. Khelif, S. Benchabane, L. Robert, V. Laude, Experimental observation of locally-resonant and Bragg band gaps for surface guided waves in a phononic crystal of pillars. Physical Review B 83 (2011), 104201. doi:10.1103/PhysRevB.83.104201.

[11] A. Bretagne, B. Venzac, V. Leroy, A. Tourin, Bragg and hybridization gaps in bubble phononic crystals. AIP Conference Proceedings 1433 (2012), 317. doi:10.1063/1.3703196.

[12] N. Kaina, M Fink, G. Lerosey, Composite media mixing Bragg and local resonances for highly attenuating and broad bandgaps. Scientific Reports 3 (2013), 3240. doi:10.1038/srep03240.

[13] Y. Chen, L. Wang, Periodic co-continuous acoustic metamaterials with overlapping resonant and Bragg band gaps. Applied Physics Letters 105 (2014), 191907. doi:10.1063/1.4902129.

[14] B. Yuan, V. F. Humphrey, J. Wen, X. Wen, On the coupling of resonance and Bragg scattering effects in three-dimensional local resonant sonic materials. Ultrasonics 53 (2013), 1332-1343. doi:10.1016/j.ultras.2013.03.019.

[15] L. Liu, M. I. Hussein, Wave motion in periodic flexural beams and characterization of the transition between Bragg scattering and local resonance. Journal of Applied Mechanics 79 (1) (2011), 011003. doi:10.1115/1.4004592.

[16] L. Raghavan, A. S. Phani, Local resonance bandgaps in periodic media: Theory and experiment. Journal of Acoustical Society of America 134 (3) (2013), 1950. doi:10.1121/1.4817894.

[17] R. Chen, T. Wu, Vibration reduction in a periodic truss beam carrying locally resonant oscillators. Journal of Vibration and Control 05 (13) (2014), 1077546314528020. doi:10.1177/1077546314528929.

[18] M. Y. Wang, X. Wang, Frequency band structure of locally resonant periodic flexural beams suspended with force-moment resonators. Journal of Physics D: Applied Physics 46 (2013), 255502. doi:10.1088/0022-3727/46/25/255502.

[19] Y. Xiao, J. Wen, X. Wen, Broadband locally resonant beams containing multiple periodic arrays of attached resonators. Physics Letters A 376 (2012), 1384 – 1390. doi:10.1016/j.physleta.2012.02.059

[20] Y. Xiao, J. Wen, D. Yu, X. Wen, Flexural wave propagation in beams with periodically attached vibration absorbers: Band-gap behavior and band formation mechanisms. Journal of Sound and Vibration 332 (2013), 867-893. doi:10.1016/j.jsv.2012.09.035.

[21] J. S. Chen, C. T. Sun, Dynamic behavior of a sandwich beam with internal resonators. Journal of Sandwich Structures and Materials 13 (4) (2010), $391 - 408$. doi:10.1177/1099636210391124.

[22] J. R. Vinson, Sandwich Structures, Applied Mechanics Reviews 54 (3) (2001), $201 - 214$. doi:10.1115/1.3097295.

[23] S. Abrate, B. Castanié, Y. D. S. Rajapakse, Dynamic failure of composite and sandwich structures, Series Solid Mechanics and Its Applications, Volume 192, Springer Science + Business Media, Dordrecht, 2013. doi:10.1007/978-94-007-5329-7.

[24] J. S. Chen, B. Sharma, C. T. Sun, Dynamic behavior of sandwich structure containing spring-mass resonators. Composite Structures 93 (2011), $2120 - 2125$. doi:10.1016/j.compstruct.2011.02.007.

[25] D. J. Mead, A new method of analyzing wave propagation in periodic structures; applications to periodic Timoshenko beams and stiffened plates. Journal of Sound and Vibration 104 (1) (1986), $9 - 27$. doi: 10.1016/S0022-460X(86)80128-6.

[26] L Liu, K. Bhattacharya, Wave propagation in a sandwich structure. International Journal of Solids and Structures 46 (2009), $3290 - 3300$. doi:10.1016/j.ijsolstr.2009.04.023.

[27] J. F. Doyle, Wave propagation in structures, Springer, New York, 1998. doi: 10.1007/978-1-4612-1832-6.

[28] D. J. Mead, Wave propagation and natural modes in periodic systems: II. Multi-coupled systems, with and without damping, Journal of Sound and Vibration 40 (1975) 19-39. doi: 10.1016/S0022-460X(75)80228-8